\documentstyle[aps,twocolumn,epsf]{revtex}
\begin{document}
\title{Ordering effect of Coulomb interaction in ballistic double-ring systems}
\author{C.~M.~Canali$^1$, W.~Stephan$^2$, L.~Y.~Gorelik$^1$,
R.~I.~Shekhter$^1$ and M.~Jonson$^1$}
\address{$^1$Dept. of Applied Physics, Chalmers University of Technology and
G\"oteborg University, S-412 96 G\"oteborg, Sweden\\
$^2$Max-Planck-Institut f\"ur Physik komplexer Systeme,
Bayreuther Str. 40, D-01187 Dresden, Germany
\\[3pt]
( Received 2 May 1997)
\\ \medskip}
\author{\small\parbox{14cm}
{\small
We study
a model of two concentric onedimensional rings with
incommensurate areas $A_1$ and $A_2$, in a constant
magnetic field. The two rings are
coupled by a nonhomogeneous inter-ring tunneling amplitude,
which makes the one-particle spectrum chaotic.
For noninteracting particles the energy of the many-body ground state and
the first excited state exhibit random fluctuations
characterized by the Wigner-Dyson statistics.
In contrast, we show that the electron-electron
interaction orders
the magnetic field dependence of these quantities, forcing them to
become periodic functions, with period $ \propto 1/(A_1 + A_2)$.
In such a strongly correlated system the only possible source of 
disorder comes from
charge fluctuations, which can be controlled by a tunable
inter-ring gate voltage.
%
\\[3pt] Keywords: A nanostructures, D. electron-electron interaction,
D. order-disorder effects, D. tunneling}}
\address{}\maketitle
An outstanding issue in the physics of mesoscopic systems
is the interplay between disorder and chaos,
and electron-electron (e-e) interaction.
Experimentally this problem is relevant
for the anomalously large persistent
currents in disordered metallic rings\cite{pcexp},
magnetoconductance fluctuations through ballistic quantum dots\cite{marcus},
Coulomb peak height fluctuations\cite{marcus2}
and peak spacings fluctuations\cite{sivan} in disordered quantum dots.
Interaction is expected to 
play a crucial role in few-electron
systems of low dimensionality; 
the way in which it affects
quantum interference in phase coherent mesoscopic systems
is presently under intense investigation.

With this motivation in mind, 
we consider here a system of
two concentric ballistic rings with incommensurate 
areas $A_1$ and $A_2$, in  a perpendicular constant magnetic
field $B$.
The phase accumulated going around each ring is
$\varphi_{p} =  B A_{p}/\phi_0,\  p=1,2$, 
where $\phi_0 = c h/e$ is the
flux quantum. We will assume that the two rings are
coupled by a {\it site-dependent} tunneling matrix
element. A particle will pick up random phases during its complicated
motion.
Such a breaking of the translational invariance can make
the system nonintegrable and the one-particle spectrum chaotic, 
even in absence of e-e interaction.

Our aim is to investigate how strong interactions
affect the energy fluctuations of the many-body ground-state and 
the first excited states as
a function of the applied magnetic field. We will show that 
the interaction {\it orders} the flux dependence of the lowest
energy levels, which
in the noninteracting case are random functions described by
the Wigner-Dyson statistics.
In contrast to the case of a clean single-channel ring\cite{loss},
in the two-ring geometry interaction plays an essential
role and modifies qualitatively the period and the amplitude of 
the persistent currents.

We choose
a lattice model of two concentric one-dimensional rings 
with ${\cal N}$ sites per ring,
described by
the following second quantized Hamiltonian
\begin{eqnarray}
{\cal H} = &&
-t\sum_{i,p}
e^{i2\pi \varphi_p/{\cal N}}\,c^{\dagger}_{i,p} 
c^{\phantom\dagger}_{i+1,p} 
-\sum_{i}t_{12}(i)c^{\dagger}_{i,1}c^{\phantom\dagger}_{i,2}+\;	H.c.
\nonumber\\&&
 + \sum_{i,j,p,p'} V(|{\bf r}_{i,p} - {\bf r}_{j,p'}|)\,n_{i,p}n_{j,p'}
\end{eqnarray}
The operators $c^{\dagger}_{i,p}$ and $c^{\phantom\dagger}_{i,p}$
create and destroy a spinless electron at site $i$ of ring $p$.
($i,j=1,\dots,{\cal N}$ are site indexes 
and $p, p' = 1,2$ are ring indexes).
The term proportional to $t$ describes the {\it intra-ring} hopping,
affected by the magnetic field.
The site-dependent amplitude $t_{12}(i)$
represents {\it inter-ring} tunneling.
The last term represents the 
translationally invariant e-e interaction. 

We first discuss the case of noninteracting particles.
If the two rings are decoupled ($t_{12}(i)=0$)
the single-particle spectrum
is given by 
$\epsilon_p(k_i) = -2 t\cos[(k_i + 2\pi \varphi_p/{\cal N})] =
-2 t\cos[(k_i + 2\pi H\,a_p/{\cal N})]$, with 
$k_i =2\pi\, i/{\cal N},\  i=0,\pm 1,\dots \pm {\cal N}/2$.
Here we have introduced two independent, dimensionless parameters,  
$H$ (magnetic field) and $a_p$ (area of ring $p$). 
When $H=1$ and $a_p=1$ the flux enclosed by ring $p$ is $\phi_0$.
Throughout this paper we will consider two incommensurate areas
$a_1 = [\sqrt(5)-1]/2\approx 0.618$ and $a_2=1.5$. In that case
the dependence of the single particle spectrum on $H$ 
will be effectively random,
with very complicated level crossings.
Suppose now that the two rings are coupled by
a tunneling amplitude $t_{12}(i)$,
small compared to the mean level spacing for each
single ring, $\Delta= 8\, t/{\cal N}$.
To first order the tunneling will couple states
of different rings.
If $t_{12}(i)$ is also site independent,
the spectrum will hardly be modified and, in particular,
level crossings will not disappear
since they always involve
levels of different momenta which are not coupled
by a homogeneous matrix element. However, if $t_{12}(i)$ is not
homogeneous, states of different momenta can be coupled, causing
the lifting of the degeneracies. Such level repulsion is
one of the essential characteristics of the Wigner-Dyson statistics.
The single-particle energies can be made interact more strongly by increasing
the magnitude of $t_{12}(i)$. We expect that when the level
splitting becomes of the order of $\Delta$,
a fully chaotic spectrum will develop. 

\begin{figure}
\vspace{0.5truecm}
\epsfxsize=4.0truecm \epsffile{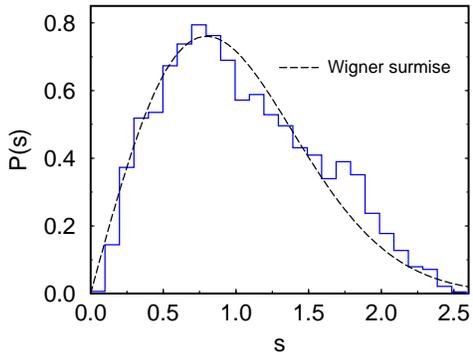}
\vspace{0.3truecm}
\caption{Probability distribution, $P(s)$, of the one-particle
energy-level spacings
for a system of two rings,
coupled by
the tunneling amplitude
$t_{12}(i) = -t/2*\cos ^2(3\pi i/2{\cal N})$.
Both rings have
11 sites; their area is
$a_1= [\protect \sqrt(5)-1]/2\approx 0.618$
and $a_2 = 1.5$ respectively.
The probability
density $P(s)$ is obtained by performing a spectral average over the energies
and an average over the flux. The dashed line represents the Wigner surmise
plotted as a comparison.}
\label{fig1}
\end{figure}

In order to show
that the one-particle spectrum is chaotic
we study
its level statistics. Statistical properties are obtained by performing
a spectral average over many energy levels for fixed $H$ and a subsequent 
average over $H$. We consider
the probability distribution of the nearest-neighbor level spacings, $P(s)$,
which we compute numerically by diagonalizing exactly a system of two rings,
both having ${\cal N} =11$ sites, coupled by 
$t_{12}(i) = -t/2*\cos ^2(3\pi i/2{\cal N})$.
For each value of $H$ the corresponding energy spectrum
is first unfolded to ensure that the mean level spacing is unity\cite{note2}.
The distribution $P(s)$ is then computed and finally averaged over many
values of $H$. In Fig.~1 we plot $P(s)$ averaged over 500 values of
$H$ between 0 and 5. The numerical curve is in good agreement with
the Wigner surmise of the orthogonal ensemble\cite{berry} characterizing the
Wigner-Dyson statistics. 

Now we might ask what are the consequences
of such a spectrum on the properties of the noninteracting many-body system.
The $H$ dependence of an experimentally relevant
quantity such as the total magnetic moment, 
$m(H) = {\partial E^N_0(H) / \partial H}$,
where $E^N_0$ is the ground-state energy for a $N$ particle system,
is just a very
irregular function of $H$, see Fig~2(a). However
the chaotic nature of the one-particle spectrum
reveals itself in the statistical fluctuations of the many-body 
ground-state and 
quasiparticle excitations as function of $H$ and $N$.
We are interested in two quantities: 
a) The distribution
of the first excitation energy above the Fermi energy, 
$\delta E^N = E^N_1 - E^N_0$;
b) The distribution of ground-state inverse compressibility: 
$\kappa^N = \partial ^2 E^N_0/\partial N^2 \sim E^{N+1}_0-2E^{N}_0 + E^{N-1}_0$.
Here $E^N_1$
is the energy of the first excited state above the Fermi energy.
It is obvious that for noninteracting
particles at zero temperature $\delta E^N$ and $\kappa^N$ are the same and 
they are equal to
$\epsilon_{N+1} -\epsilon_{N}$, where $\epsilon_{N}$ is the $N^{\rm th}$
single-particle energy. Thus the fluctuations of $\delta E$ and $\kappa$,
averaged over the Fermi energy (namely over $N$) and $H$, scale with
the $\Delta$ and their distribution is given by the
same $P(s)$ of the one-particle spectrum which, for our coupled-ring system, 
is given by the Wigner surmise.

How is this complex scenario modified by strong e-e
interactions? To answer this question we have resorted to exact
numerical calculations on small systems, using the Lanczos algorithm
to compute the fluctuations of many-body ground state and the 
first low-lying excitations as a function of $E_F$ and $H$.
We have considered systems with ${\cal N}=11$ sites per ring and
${N} =4,5,6,7,8$ particles.
The number of electrons in each ring, ${N}_p$, is
not a good quantum number if the rings are coupled.
The expectation values $\langle{N}_p\rangle$ will however satisfy 
the constraint $\langle {N}_1\rangle +\langle  {N}_2\rangle= {N}$.
We will first discuss the case of a short range
interaction, coupling two nearest-neighbor 
sites $(i,p)$ and $(i+1,p)$ 
in the same ring with matrix element $V_{11}$,
and opposite sites $(i,1)$ and $(i,2)$ in different rings
with matrix element $V_{12}$.
The relevant energy parameter, indicating the relative strength
of the interaction, is $V_{p p'}/t$.
It is interesting to see how $m(H)$ is modified.
A plot of $m(H)$ vs $H$ for a system of $N= 8$ particles
interacting with
$V_{11} =6.5\,t,\ \ V_{12} = 6.3\,t$
is shown in Fig.~2a. 
The effect of the interaction is visibly spectacular.
We can see that the aperiodic
$H$-dependence of $m(H)$ of the noninteracting case is replaced 
by a perfectly {\it periodic} function, oscillating with period
$H_0 = 1/(a_1 + a_2)$, exactly as we would have for a single ring with area 
$a_{\rm tot}= (a_1 + a_2)$. The periodic $m(H)$ develops smoothly 
with increasing interaction and eventually it becomes
independent of the interaction strength.
The amplitude of $m(H)$ in the strongly interacting regime
is approximately one half of the amplitude of the noninteracting case.
\begin{figure}
\epsfxsize=4.7truecm \epsffile{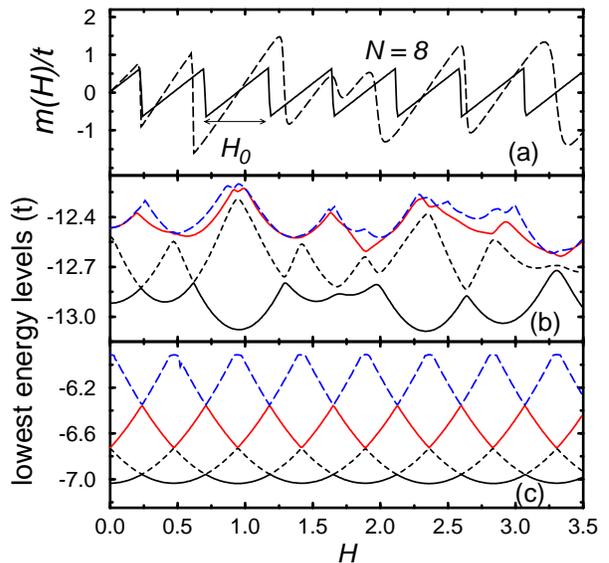}
\vspace{0.5truecm}
\caption{(a) Magnetic moment $m(H)$ vs $H$, and
(b), (c) four lowest many-body energies
vs $H$ for the two-ring system of Fig.~1.
The total number of particles is $N = 8$.
(a) The periodic solid line is the interacting case. 
The period is $H_0= 1/(a_1 + a_2)$; the irregular
dashed line is the noninteracting case.
(b) Noninteracting case.
(c) Interacting case.}
\label{fig2}
\end{figure}
Moreover, not only is the ground-state energy
a periodic function of $H$, but so are the first low-lying excited states,
as we show in Fig.~2(c).
In particular the low energy sector of the many-body
spectrum is identical to the single-particle spectrum in one ring.
Similar results are obtained for other
values of even ${N}$. 
By computing the ground-state wavefunction we can also evaluate the particle
density in each ring.
The calculation shows that in the strongly interacting case
$\langle N_1\rangle \approx\langle N_2\rangle \approx {N}/2$
for any value of $H$. 
For odd ${N}$
the interaction again forces
the system to develop a periodic low-energy spectrum
with the same period, $1/(a_1 + a_2)$.
The main difference with respect to even ${N}$ is that
the $H$ dependence of the spectrum is shifted by 
$H_0/2$. 

Strong intra-ring interaction is known to generate Wigner-crystal-like
ground states in each single ring\cite{shultz}.
It has also been shown that inter-ring interaction can cause a nondissipative
Coulomb drag\cite{rojo} that affects qualitatively the persistent 
currents in each ring\cite{ulloa}.
Our exact results suggest that strong inter-ring interaction
glues together the two stiff Wigner crystals created by the intra-ring
interaction in each ring, resulting in a 
rigid structure that rotates like a single solid body
under the effect of the magnetic field.
Based on this picture, we can easily see why the
oscillation period of the total magnetic moment of such a
structure 
is $1/(a_1 + a_2)$. Indeed the Hamiltonian of the two frozen crystals
glued together,
each containing $N/2$ particles,
is found to be

\begin{equation}
\label{ham2wc}
{\cal H} = 
{\hbar^2\over 2\,I} \Bigl[-i\,{\partial\over\partial\theta} - {N \over 2} 
H {(a_1 + a_2)}\Bigr]
\end{equation}
where $-i\,\hbar {\partial \over \partial \theta}$ 
is the canonical momentum conjugated to $\theta$,
the rotation angle of the solid around an axis passing through
the center of the rings. $I= m_e\, {N \over 2\pi} (A_1 + A_2)$ 
is the moment of inertia. In zero flux, the eigenstates
$\Psi_n(\theta) =\exp (i\,n\,\theta)$ must
satisfy the boundary condition 
$\Psi_n(\theta) = \Psi_n\bigl(\theta + 2\pi/(N/2)\bigr)$ due to the ordered
crystal structure. This immediately implies the announced $1/(a_1 + a_2)$
periodicity as a function of $H$.

We have also studied the case of an unscreened
long range interaction coupling all the sites with matrix element
$V(|{\bf r}_{i,p} - {\bf r}_{j,p'}|) = V_c R_1/|{\bf r}_{i,p} -{\bf r}_{j,p'}|= 
V_c\,\sqrt{a_1}/ \Bigl [a_p + a_{p'} -2\sqrt{a_p\; a_{p'}} 
\cos \bigl(2\pi(i-j)/{\cal N}\bigr)\Bigr ]^{1/2}$.
The dimensionless interaction constant is 
$\alpha= V_c R_1/\hbar v_{\rm F} = (V_c/t)[{\cal N}/ 4\pi \sin (k_{\rm F})]$.
In this case we have to face first a
new occurrence: 
in the strongly interacting regime ($\alpha \approx 5$) 
where spectrum ordering should set in,
the electronic configuration corresponding to the minimum energy 
is always
the one with the inner ring empty and all the particles located
in the outer ring. The corresponding low energy sector of the spectrum is
obviously a periodic function of $H$ with period $1/a_2$. 
To by-pass this unwanted situation
we can 
add a term in the
Hamiltonian in the form of a potential
difference between the rings, e.g. provided by a gate voltage:
$H_{g} = \sum_{i,p} \epsilon_{p}n_{i,p}$.
Typically a voltage
difference $V_g = \epsilon_1 -  \epsilon_2 \approx -V_c/2$ ensures
an almost equal population in the two rings.
In Fig.~3 we plot $m(H)$ for $N= 6$ particles interacting
with a Coulomb force of intensity $V_c = 10\, t$, for two values of the voltage 
$V_g$.
Let us consider first zero inter-ring tunneling (solid line)\cite{note1}.
We can see that, for $V_g =-0.75\,V_c$, $m(H)$ is perfectly periodic 
with the expected period $1/(a_1 + a_2)$.
On the other hand imperfections and small glitches in the periodic 
pattern appear
for $V_g =-0.5\,V_c$.
In fact the calculation of the ring occupancy shows
that while for $V_g =-0.75\,V_c$ the electronic configuration
$(N_1 =3, N_2=3)$ is stable for all 
values of H, when $V_g =0.5\,V_c$ the
two configurations $(N_1 =3, N_2=3)$ and $(N_1 =2, N_2=4)$ alternate in the
ground state. 
The aperiodic jumps in Fig.~3(a) correspond to a switch from
one configuration to another as $H$ varies.
It is instructive to look at the $H$ dependence of the
lowest four energy levels, plotted in Fig.~4, and their corresponding
ring density.
Fig.~4(a) shows that
when $V_g= -0.5\,V_c$ these four levels are often two by two degenerate,
the degeneracy being caused
by the presence of the two electronic configurations
$(N_1 =3, N_2=3)$ and $(N_1 =2, N_2=4)$
mentioned above. The
gate voltage is not yet strong enough to enforce a fixed electronic
configuration in the two rings.
As a result the levels are not completely
periodic and they are sensitive to a inter-ring tunneling-- see dashed line of
Fig~3(a). 
However if we increase $V_g$ up
to $0.75\,V_c$, the configuration $(N_1 =3, N_2=3)$ is locked
in the first three states. 
It is only starting from the fourth level
that charge redistribution, with the appearance of the
$(N_1 =2, N_2=4)$ configuration at some values of $H$, is again possible. 
The important consequence is the 
ordering of the first three levels, which is robust against a nonhomogeneous
tunneling between the rings, as shown in Fig.~4(c) and
Fig.~3(b). 
\begin{figure}
\vspace{0.25truecm}
\epsfxsize=4.5truecm \epsffile{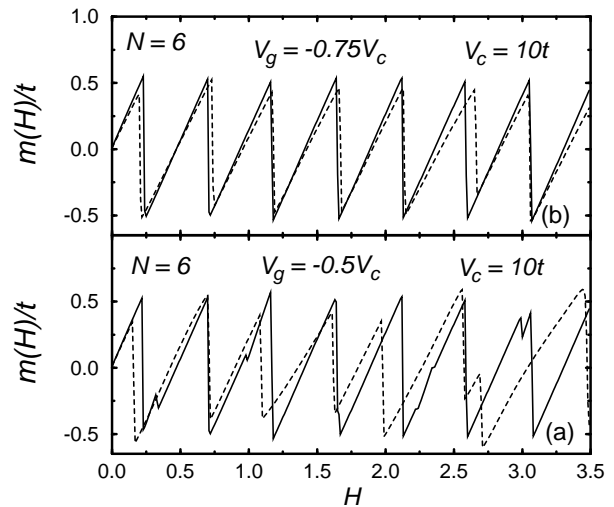}
\vspace{0.25truecm}
\caption{Magnetic moment vs $H$ for the system
of Fig.~1. There are ${N}=6$
particles interacting with a long-range
Coulomb potential of intensity $V_c = 10\,t$.
The {\it periodic} solid line represents the case
of $t_{12}(i)= 0$.
The oscillating period is $1/(a_1 + a_2)$.
The dashed line represents the case of
$t_{12}(i) = -t/2\,\cos^2(2\pi i/{\cal N})$. A voltage difference
$V_g$ 
is applied between the two rings.
a) $V_g = -0.5\, V_c$. (b) $V_g = -0.75\, V_c$.}
\label{fig3}
\end{figure}

\begin{figure}
\epsfxsize=4.7truecm \epsffile{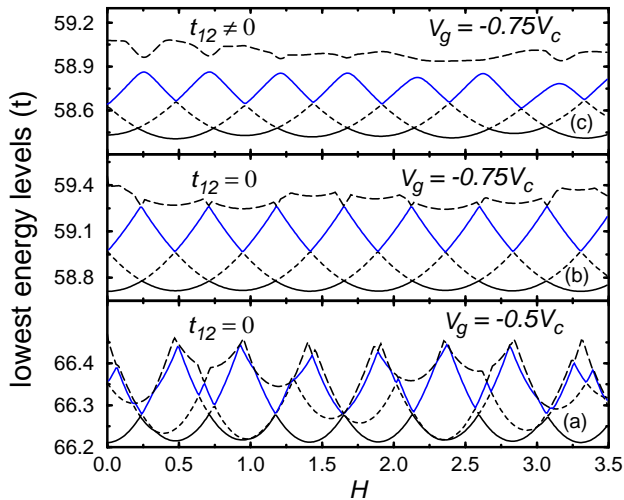}
\hspace{-0.5truecm}
\caption{Energy vs $H$ of the ground state and the first three excited states
for the system of Fig.~3.
(a) The two rings are decoupled [$t_{12}=0$]
and the voltage difference $V_g$ between the rings
is -0.5 of the Coulomb interaction intensity, $V_c =10t$.
(b) $t_{12}=0$ and $V_g= -0.75\, V_c$.
(c) $t_{12}(i) = -t/2\,\cos^2(2\pi i/{\cal N})$
and  $V_g= -0.75\, V_c$.}
\label{fig4}
\end{figure}
\noindent 
The example of 6 particles is representative of what should happen
in a real mesoscopic system with $N_1\sim N_2 \sim 50$, where gate
and flux-induced-charge redistribution between the rings amounts to
small changes in their electron concentration.
If the gate voltage is strong enough to enforce an almost equal population 
and prevent charge rearrangement between the rings as $H$ varies, 
then the e-e Coulomb interaction
can fully develop the ordering effect that we have discussed.
Disorder in this strongly correlated system can only be induced by
charge redistribution, which is controllable
by tuning $V_g$. 
Our calculations show that spectrum ordering is already well developed
for $\alpha \approx 5$. Experimentally, GaAs-Al/GaAs heterostructures 
with $\alpha > 1$ are normally available.
Thus the effect that we have found might be detected
in a system of two semiconductor rings. 

We now go back to the problem of the fluctuations 
of $\delta E$ and $\kappa$ and study their fate 
in the strongly interacting regime.
Because of the periodicity in $H$, the energies of the 
ground-state and the first excited
state in the interval $0< H < H_0/2$
can be approximately parametrized by :
\begin{eqnarray}
E_0^N(H) &=& {V_c\over C}N^2 + \Delta_N\; H^2\;, \\
E_1^N(H) &=& {V_c\over C}N^2 + \Delta_N\; (H_0 -H)^2\;,
\end{eqnarray}
for even $N$, while the expression for odd $N$ are obtained
shifting $H$ by $H_0/2$.
Here $\Delta_N \approx \Delta (N/{\cal N})$ is a slowly increasing 
function of $N$ and $C$ is a constant.
Thus for fixed $N$ and $H$ we obtain,  
\begin{eqnarray}
\delta E^N(H) &=& \Delta_N (H_0^2 - 2H_0H)\;, \\
\kappa^N(H) &=& 2{V_c\over C} - \Delta_N (H_0^2/2 - 2H_0H)\;
\end{eqnarray}
Notice that in the interacting case the fluctuations of $\kappa$ scale with
the Coulomb energy.
To obtain the probability distribution of $\delta E$ and $\kappa$,
for each value of $H$ we first need to unfold the sequence of $\delta E^N$ and
$\kappa^N$ obtained upon varying $N$, to take into account the monotic change
of $\Delta_N$. Once this is done, no fluctuations around the average
are left and the distributions of $\delta E$ and $\kappa$ are just
delta functions centered at $s= 1$ in rescaled units. Since this result
is valid for any value of $H$, the average over $H$ does not add anything.
This is in sharp contrast to the case of noninteracting particles,
where we saw that the two distributions were given by the
Wigner surmise.

In conclusion, we have shown that  e-e interaction
orders the flux dependence of the low energy spectrum 
of a two-ring system, suppressing the irregular behavior of the
free particle case. The fluctuations of the
ground-state and first excited state are no longer described by the
Wigner-Dyson statistics.
When strong e-e correlations are present,
disorder can only be induced by charge redistribution controllable
through an external gate.

We thank I.~V.~Krive for useful discussions. 
Financial support from NFR and TFR is
gratefully acknowledged.



\begin{references}
\bibitem{note2}
        The unfolding is done by choosing a new 
        energy variable ${\cal E} = {\cal E}(\epsilon)$,
        through the relation $d{\cal E}/d\epsilon = \rho(\epsilon)$.
        $\rho(\epsilon)$ is the average density of states in the original
        variable $\epsilon$.
\bibitem{note1}
	Among all the configurations $(N_1, N_2)$,
	satisfying the condition $N_1 + N_2= N$,
        the one of minimum energy is taken. Such
	$(N_1, N_2)$ depends, in general, on $H$.
\end{references}
\end{document}